\documentclass[aps,prl,floatfix]{revtex4}
\usepackage{graphicx}
\usepackage{color}
\usepackage{xcolor}
\usepackage{mathrsfs}
\newcommand{\ket}[1]{{|#1\rangle}}
\newcommand{\bra}[1]{{\langle#1|}}
\begin{document}

\title{Coherence, Entanglement and Complementarity in Mixed Classical Light
\vspace{-2mm}}

\author{Asma Al-Qasimi}
\affiliation{Department of Physics and Astronomy and Center for Coherence and Quantum Optics, University of Rochester, Rochester, NY 14627, USA}

\vspace{-2mm}
\begin{abstract} \noindent

%Polarization coherence in the Young's Interference Experiment is directly related to the strength of the visibility of fringes. 
The classical analogues of quantum correlations have been the subject of considerable interest in the past few years; however, all of these investigations consider the classical analogue of pure quantum mechanical states only. In this work, we study {\em mixed} classical light states and derive relations between the polarization coherence of the field and the (classical) entanglement between its degrees of freedom. We show, for a specific model of mixed states, where the purity is determined by a single parameter, how the coherence shared between polarization and entanglement shrinks as the level of purity decreases. The sum of the square of polarization and entanglement remains constant, a behaviour consistent with previous results \cite{pureCP} of pure states, even though the constant for the mixed state case is now smaller in value.

\end{abstract}

\maketitle

\section{Introduction}

In recent years, various authors have advanced a general description of light using linear algebraic notions familiar from quantum mechanics to consider all its degrees of freedom (DOFs) \cite{Sprew1998,Sprew2001,QianOL,Eberly2015,Eberly2016,DeZela,AD2019,AAi}. This more convenient description of light allows for the study of coherences that were not accessible using the traditional description \cite{Eberly2015}. Specifically, the correlations between different DOFs are easier to consider: depending on context, if one wants to focus on specific DOFs, one either \emph{projects} the total state onto a specific state of the DOF to be ignored, or one \emph{traces over} the unwanted DOF \cite{Eberly2016,EberlyScripta2016}.

This new formalism for describing classical light shows the connection between the entanglement of the various DOFs with specific coherences, whether it is the traditional polarization coherence, which has to do with the collection of wave vibrations in a specific direction, or with the \emph{mode} polarization coherence, which touches on the spatial properties of light and determines the sharpness of the visibility fringes \cite{Abouraddy}. In this context, entanglement is the measure of how unfactorizable a wavefunction is in terms of functions of its DOFs; since the wavefunctions are classical electromagnetic fields rather than quantum states, the physical content is different.

Generally speaking, previous investigations \cite{pureCP,QianOL} of complementarity in classical optics have concentrated on deterministic, as opposed to stochastic fields, analogous to pure states in quantum mechanics. A manifestation of being in a pure state is to have the maximum value for the \emph{interference} terms for the intensity. For example, if we consider polarization coherence, this implies the \emph{equality} in the Cauchy-Schwarz inequality for averages of field components $ \langle E_{x}^{*} E_{x} \rangle \langle E_{y}^{*} E_{y} \rangle  \geq | \langle E_{x}^{*} E_{y} \rangle|^2$ (where the brackets represent the statistical averaging of the various perpendicular components of the electric field, which is taken to be travelling in the $z$ direction here). 

In this paper, we study the case of non-deterministic (in the classical sense) mixed states, where only the \emph{inequality} (without the equality) part of the Cauchy-Schwarz relation holds, making it impossible to write the state of the field as a linear combination of the various DOFs. Instead, a matrix (analogous to the density matrix in quantum mechanics) becomes necessary for the field description, where the diagonal elements describe the intensity due to a specific wave, while the off-diagonal elements correspond to interference between the various waves/states. We study mixed states of light, generalizing the previously discovered relationship between entanglement and polarization coherence \cite{pureCP}. We also consider a specific model, with one parameter that quantifies the purity, and provide a visual representation that demonstrates how this relationship changes as the electric field state becomes more mixed.

\section{Field description, notation and mathematical tools}

For the monochromatic fields that we are considering, we call the directions' DOFs that define the vibration the \emph{spin} DOFs, as the term \emph{polarization} can be ambiguous. Each spin direction has an amplitude DOF associated with it, which is a function of position, which we represent in the cartesian coordinates $(x,y,z)$. These amplitudes determine the significance of each direction in the field description, which ultimately defines the polarization of the field \cite{BornWolf,Hecht}. We assume the field is paraxial; it does not have a component in the direction of propagation; i.e., the $z$ direction.

We can thus choose to represent this vector function using appropriate orthonormal mode functions, i.e.
\begin{equation}
{\bf E}(x,y,z) = \sum_{i=1,2} \sum_{k} A_{ik} \hat{\bf s}_i Q_k (x,y,z),
\label{Ee}
\end{equation}
where $A_{ik}$ represents constants, $\hat{\bf s}_i $ is a unit vector specifying the spin direction, and $\{Q_k (x,y,z)\}$ represents the position-dependent modes that describe the spatial variation of the field.  Each of the DOFs (spin and position) form a separate vector space \cite{Eberly2015, Eberly2016}; the field is a vector in a vector space formed by a tensor product of these two subspaces. One can adopt the Dirac notation from quantum mechanics to write (\ref{Ee}) as follows \cite{Eberly2015, Eberly2016}:
\begin{equation}
\ket{\bf E} = \sum_{i k} A_{ik} \ket{s_i} \otimes \ket{Q_k}.
\label{ketE}
\end{equation}

\noindent We unit-normalize the field by writing
%%%%%%%%%%%%% 
\begin{equation}
\ket{\bf e} = \sum_{i k} b_{ik} \ket{s_i} \otimes \ket{Q_k},
\label{normalizedE}
\end{equation}
where $\sum_{i k} |b_{ik}|^2=1$, $\langle{s_i}|{s_j}\rangle=\delta_{ij}$, and $\langle{Q_x}|{Q_l}\rangle=\delta_{xl}$, and with ${\bf e} ={\bf E}/\sqrt{I}$, where $I={\bf E}^*\cdot{\bf E}$ is the intensity.

Before we discuss the appropriate representation for mixed states, the focus of this work, we consider the matrix representation of the pure state in (\ref{normalizedE}). This matrix carries the interference terms (between the states involved in the field description). For pure states, such as the one in (\ref{normalizedE}), it is simply the state matrix $\hat{\cal{W}}=\ket{\bf e}\bra{\bf e}$. In this case, the relation $\hat{\cal{W}}^2=\hat{\cal{W}}$ holds. A mixed state also has a state matrix representation, but this condition no longer holds, and, therefore, it is not possible to represent it by a description such as the one given in (\ref{normalizedE}).

%In our study, we will consider a situation that deals with a state with two-state DOFs, so we look at the Young Interference experiment, depicted in figure \ref{fig:figure1}. This is a two-path-two-polarization scenario, since as the field (\ref{normalizedE}) passes through the two pinholes, this forces the selection of spatial modes (the spherical waves emanating from the two pinholes), depending on two positions (location of the pinholes), leading to the following state, just as the field passes the screen:

\begin{figure}
  \includegraphics[width=0.5\linewidth]{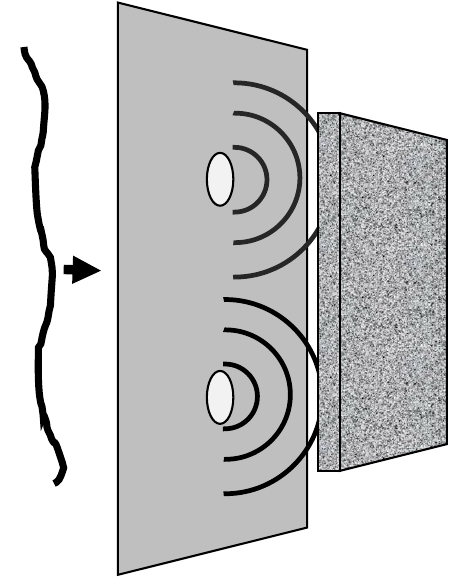}
  \caption{{\bf The Electromagnetic field in the Young Experiment with a randomizer/decohering system.} The field is incident on the two-pinhole screen from the left.  The pinhole projects the field into two distinct spatial modes (i.e. the radiation field diffracted by each of the pinholes), while the spin degree of freedom is unaltered. The field then passes through a random linear optical system (shown {\em schematically} as a slab) which performs some random unitary transformation to the light (e.g. by changing phases and mixing the degrees of freedom using beam-spliters, wave-plates etc.)}
  \label{fig:figure1}
\end{figure}

In our study, we consider the Young Interference experiment, depicted in figure \ref{fig:figure1}. It is a two-path-two-polarization scenario, since as the field (\ref{normalizedE}) passes through the two pinholes, this forces the selection of spatial modes (the spherical waves emanating from the two pinholes), depending on two positions (location of the pinholes), leading to the following state, just as the field passes the screen:
\begin{equation}
\ket{\bf e} = b_{11} \ket{s_1 {Q_1}} +b_{12} \ket{s_1 {Q_2}} +b_{21} \ket{s_2 {Q_1}} +b_{22} \ket{s_2 {Q_2}},
%\ket{\bf e} = b_{11} \ket{s_1 {\bf r_1}} +b_{12} \ket{s_1 {\bf r_2}} +b_{21} \ket{s_2 {\bf r_1}} +b_{22} \ket{s_2 {\bf r_2}}.
\label{E2}
\end{equation}

\noindent where $\ket{Q_i}$ corresponds to the spatial mode coming out of the $i^{th}$ pinhole.

%++++++++++++++++++++++++++
%
%++++++++++++++++++++++++++

The focus of this work is on two properties of light: polarization coherence (quantified by the degree of polarization, \emph{P}, and usually describes how close to one direction the electric field vibrates in) of the light field and the entanglement between its DOFs. Although in the Young experiment example to be discussed here, \emph{P} refers to the spin DOF, in general, it does not have to.

The degree of polarization \emph{P} is defined by using the standard definition (See \cite{WolfsLittleBook}, page 163):
\begin{equation}
P=\sqrt{1-4 \mbox{Det}\{\hat{J}\}},
\label{PJ}
\end{equation}
\noindent where $\hat{J}$ is the polarization coherence matrix (see \cite{WolfsLittleBook}, section 8.1, eq. (2)).  This matrix is obtained from the state matrix of the system by tracing out the spatial DOF.
%and its eigenvalues, $\lambda_1$ and $\lambda_2$, are real and positive, and they add up to 1 (since the state is normalized). Without  of generality, we shall assume $\lambda_1\ge\lambda_2$. 
From (\ref{normalizedE}), one can see that the $i^{th}$ spin component of $\ket{\bf e}$ is given by $ \sum_{k} b_{ik} \ket{Q_k}$, so that $J_{ij}=\langle{\bf e_i}|{\bf e_j}\rangle = \sum_{kl} \langle{Q_l}|{Q_k}\rangle b_{jk}b^{*}_{il}=\sum_{k} b_{jk}b^{*}_{ik}$. Notice how the inner products between the positional mode vectors, $\ket{Q_k}$, look very similar to Wolf's representation of the coherency matrix elements, given by the correlations between the various components of the electric field \cite{WolfsLittleBook}. The 2 x 2 coherency matrix can, therefore, be represented by:
\begin{eqnarray}
%\hat{J}&=&\mbox{Tr}_r\{(\ket{\bf e}\bra{\bf e})\}\nonumber\\
\hat{J}&=&\mbox{Tr}_r\{\hat{\cal{W}}\}\nonumber\\
&=& \sum_{i,j=1}^2 J_{ij} \ket{s_i} \bra{s_j},
%&=& \sum_{p=1}^2 \lambda_p\ket{\tilde{s}_p} \bra{\tilde{s}_p},
\end{eqnarray}

To quantify entanglement of a pair of two-dimensional systems in a pure state, one writes the projector of the state in its matrix representation, then trace out one DOF and see how much randomness is introduced into the system (by calculating the purity of the remaining DOF). The idea is that if there is entanglement in the system, by tracing out the DOF, one loses information about the correlations between the two DOFs, which results in the state of the remaining DOF becoming mixed. On the other hand, if there is no entanglement, then the purity will remain unity, as these DOFs are statistically independent of each other in this case. This same concept can be extended to the mixed state case by using the fact that mixed states have pure-state decompositions, so the idea is to find out the entanglement of each of the pure states involved in the decomposition and weight them accordingly. Moreover, since the pure-state decomposition is not unique, one has to minimize the entanglement over all possible pure-state decompositions. This idea is the basis for Wootters' remarkable quantification of mixed-state entanglement known as Concurrence \cite{Wootters}.

To study the entanglement between the two two-dimensional subspaces (position and spin), we use Wootters' Concurrence \cite{Wootters}, $C$. It can be shown for the state in (\ref{E2}) to be $C=2|b_{11}b_{22}-b_{12}b_{21}|$. Moreover, for the same state, it can be shown that the polarization coherency matrix is given by:

\begin{equation}
\hat{J}=\left(
\begin{array}{cc}
|b_{11}|^{2}+|b_{12}|^{2} & 
b_{11}b_{21}^{*} +b_{12}b_{22}^{*}\\
%%%%%%%%%%%%%%%%%%%%%%%%%%%%%
b_{21}b_{11}^{*}+b_{22}b_{12}^{*}& 
|b_{21}|^{2}+|b_{22}|^{2}\\
\end{array}
\right).
\label{Jeg}
\end{equation}

Using (\ref{Jeg}) and the definition for the degree of polarization in (\ref{PJ}), we obtain:

\begin{equation}
P=\sqrt{1-4|b_{11}b_{22}-b_{12}b_{21}|^{2}}.
\label{Peg}
\end{equation}

Combining these two expressions, we obtain:
\begin{equation}
C^2+P^2= 1
\label{CPtheorem}
\end{equation}
i.e., there is a relationship between the entanglement between the two DOFs and the degree of polarization in one of the DOFs. For pure states, it is an equality that always holds: if $P$ decreases, $C$ has to increase to compensate for that, and vice versa. This is not a new result for pure states. In \cite{pureCP},  a more general situation that looks at the various possibilities for \emph{P} in an experiment, along with the experimental results, is discussed, and the result in (\ref{CPtheorem}) is already derived. Here, however, we generalize the problem by looking at systems that are not pure, but mixed.

\section{Results and discussion}

Let us consider a simple, experimentally plausible model, which has the virtue of allowing us to calculate quantities in closed form. Suppose the pure incident field in figure \ref{fig:figure1} encounters a linear decohering system (depicted by the textured slab in the figure) causing it to become mixed, due to the two propagating fields acquiring differing phase modifications for their independent DOFs (spins and paths), thus affecting the coefficients in (\ref{E2}): 
$b_{11}\rightarrow b_{11}\exp(i\phi_0)$, $b_{12}\rightarrow b_{12}\exp(i\phi_1)$, $b_{21}\rightarrow b_{21}\exp(i\phi_2)$ and $b_{22}\rightarrow b_{22}\exp(i\phi_3)$.  On taking the ensemble average (denoted by an overbar), the state matrix $\hat{\cal{W}}=\overline{\ket{\bf e}\bra{\bf e}}$ has the following matrix representation:
\begin{equation}
\hat{\cal{W}}=\left(
\begin{array}{cccc}
|b_{11}|^2 & 
b_{11}b_{12}^\ast\overline{\exp[i(\phi_0-\phi_1)]} & 
b_{11}b_{21}^\ast\overline{\exp[i(\phi_0-\phi_2)]} & 
b_{11}b_{22}^\ast\overline{\exp[i(\phi_0-\phi_3)]}\\
%%%%%%%%%%%%%%%%%%%%%%%%%%%%%
b_{12}b_{11}^\ast\overline{\exp[i(\phi_1-\phi_0)]} & 
|b_{12}|^2 & 
b_{12}b_{21}^\ast\overline{\exp[i(\phi_1-\phi_2)]} & 
b_{12}b_{22}^\ast\overline{\exp[i(\phi_1-\phi_3)]}\\
%%%%%%%%%%%%%%%%%%%%%%%%%%%%%
b_{21}b_{11}^\ast\overline{\exp[i(\phi_2-\phi_0)]} & 
b_{21}b_{12}^\ast\overline{\exp[i(\phi_2-\phi_1)]} & 
|b_{21}|^2 & 
b_{21}b_{22}^\ast\overline{\exp[i(\phi_2-\phi_3)]}\\
%%%%%%%%%%%%%%%%%%%%%%%%%%%%%
b_{22}b_{11}^\ast\overline{\exp[i(\phi_3-\phi_0)]} & 
b_{22}b_{12}^\ast\overline{\exp[i(\phi_3-\phi_1)]} & 
b_{22}b_{21}^\ast\overline{\exp[i(\phi_3-\phi_2)]} & 
|b_{22}|^2
\end{array}
\right)
\label{star}
\end{equation}
For simplicity, we assume that the random phases have zero mean, identical variances $\overline{\phi_i\phi_i}=\sigma^2$, and Gaussian statistics, so that 
$\overline{\exp[i(\phi_i-\phi_j)]}= \exp[-\sigma^2(1-\gamma_{ij})]$, where $\overline{\phi_i\phi_j}=\sigma^2\gamma_{ij}$.  Further, let us assume that the variance $\sigma^2$ is large (so that $\exp[-\sigma^2]\ll1$), and that $\phi_0$ and $\phi_3$ are completely uncorrelated with either of $\phi_1$ and $\phi_2$, so that $\gamma_{01}=\gamma_{02}=\gamma_{31}=\gamma_{32}=0$.  We will assume that $\phi_0$ and $\phi_3$ and $\phi_1$ and $\phi_2$ are both strongly (but not perfectly) correlated with each other, so that $\overline{\phi_0\phi_3}=\sigma^2-\xi$ and $\overline{\phi_1\phi_2}=\sigma^2-\eta$, where $\xi$ and $\eta$ are parameters of order unity which can be controlled experimentally. The partially coherent state in (\ref{star}) can be realized experimentally in various ways; for example, using variable liquid-crystal retarders or electro-optic modulators in front of the two pinholes. These can be driven by partially-correlated random driving voltages (generated, for example, by linear superposition of uncorrelated noise signals). The resultant random modulation of the birefringence introduces correlated random phase differences between the vectorial components of the two beams. The coherence time of the phase fluctuations is assumed to be much shorter than the coherence time of the instant light. All of these are reasonable assumptions to make about an experiment. In these circumstances, the density matrix reduces to the following generic form \cite{xstate,NAD}
\begin{equation}
\hat{\cal{W}}_{X}=\left(
\begin{array}{cccc}
a & 0 & 0 & w \\
%&&&\\
0 & b & z & 0 \\
%&&&\\
0 & z^* & c & 0 \\
%&&&\\
w^* & 0 & 0 & d 
\end{array}
\right),
\label{rhoX}
\end{equation}
where the diagonal elements are $a=|b_{11}|^2$, $b=|b_{12}|^2$, $c=|b_{21}|^2$ and $d=|b_{22}|^2$, and the off-diagonal elements are $z=b_{12}b_{21}^\ast\exp(-\eta)$ and $w=b_{11}b_{22}^\ast\exp(-\xi)$. The matrix element $a$ corresponds to the $\ket{s_1,Q_1}\bra{ s_1,Q_1}$ element of the decohered matrix, and $w$ corresponds to the $\ket{s_1,Q_1} \bra{ s_2,Q_2}$ element of the matrix, and so on. Since $|z|\le \sqrt{b c}$ and $|w|\le \sqrt{a d}$, this matrix is always positive, as it should be for a density matrix.

In this case, the Concurrence can be shown to be \cite{NAD}
$C=2 \mbox{max}\{0,|z|-\sqrt{a d},|w|-\sqrt{b c}\}$ and the Polarization $P= \sqrt{1-4(a+b)(c+d)}$. Using the positivity condition  \footnote{To elaborate on the positivity condition used here, the density matrix is a positive definite matrix; its eigenvalues are positive, and its diagonal elements are positive. Using these two properties, we can derive conditions, in terms of the elements of the density matrix, that must be satisfied to maintain positivity. Then, we use the expressions for $C$ and $P$ (in terms of these elements) along with these conditions and, through algebraic manipulation, show that $C^2+P^2\le1$.}, we get the following upper limit on Concurrence, viz.: $C\le 2|\sqrt{ad}-\sqrt{bc}|$.
Simple algebra shows that:
\begin{eqnarray}
C^2+P^2&\le& 4 (\sqrt{ad}-\sqrt{bc})^2+1-4(a+b)(c+d)\nonumber\\
&=&1-8\sqrt{ad}\sqrt{bc}-4(a c+bd)\nonumber\\
&=& 1-4 (\sqrt{ac}+\sqrt{bd})^2\le1.
\end{eqnarray}
Hence, $C^2+P^2\le1$.  The replacement of the equality eq.(\ref{CPtheorem}) will clearly have consequences on relations such as the \emph{polarization coherence theorem} (PCT) discussed in \cite{EberlyOptica}, or its experimental manifestation \cite{DeZela2}. %\footnote{There have been other contributions in this context to note. In arXiv:quant-ph/0302075,  \emph{Phys. Rev. A} \textbf{76}, 052107 (2007), and \emph{Opt. Comm.} \textbf{283}, 827 (2010), Jakob and Bergou considered notions of duality relevant to \emph{two-particle} interference, which they analyzed in terms of single-particle visibilities and obtained results which superficially are analogous to the PCT. In the current paper,  I am dealing with interference in classical fields, which quantum mechanically is analogous to single-particle wave functions. Complemantarity  and duality specifically refers to \emph{single} particles having wave-like or particle-like properties. Hence, Jakob and Bergou's two-particle notion is of marginal relevance to the case here.}

% \footnote {Based on numerical investigations and preliminary algebraic results, the inequality $C^2+P^2\le1$ does seem to be true for all mixed states, not just the ''X''-states discussed here; this will be the topic of an upcoming publication (D. F. V. James and A. Al-Qasimi, ``Complementarity inequality for mixed states of two qubits'' {\em to be submitted}).}

To illustrate how the complementarity relationship is altered, and thus get a better feel for this relationship, we need to have a specific example which will allow us to find exact values, not in the form of an inequality. To achieve this, we study $\hat{\cal{W}}_{X}(\epsilon)$, a special case of the X-state, $\hat{\cal{W}}_{X}$, with $a$, $b=c$, $d$, and $w$ all being a function of the parameter $\epsilon$, and $z=0$. This single parameter determines how mixed the state is. To elaborate on how $\hat{\cal{W}}_{X}(\epsilon)$ is obtained, it is constructed from a maximally mixed state (proportional to the identity operator) and a pure state; i.e, $\hat{\cal{W}}_{X}(\epsilon)$ has the defined form:

\begin{equation}
\hat{\cal{W}}_{X}(\epsilon)= \frac{\epsilon}{4 }\hat{I}+(1-\epsilon)\ket{\psi}\bra{\psi},
\label{rhoMP}
\end{equation}

\noindent where $\hat{I}$ is the 4 x 4 matrix representing the identity operator, and
\begin{equation}
\ket{\psi}=\cos\theta\ket{u_a}\ket{v_a}+\sin\theta\ket{u_b}\ket{v_b},
\label{psiSch}
\end{equation} 
\noindent where the vectors labelled by $a$ are orthogonal to the ones labelled by $b$, and $\left \{ \ket{u_a},\ket{u_b} \right \}$ and $\left \{ \ket{v_a},\ket{v_b} \right \}$ represent the spin and spatial DOFs, respectively, and are related to the original DOFs described in (\ref{E2}) as follows: $\ket{u_a}\ket{v_a}=\ket{s_1}\ket{Q_1}$ and $\ket{u_b}\ket{v_b}=\ket{s_2}\ket{Q_2}$. The model in eq. (\ref{rhoMP}) is a simple and intuitive way of adding uncorrelated noise to a pure state. It is constructed in a very similar fashion as the Werner states \cite{Werner} in the quantum entanglement literature: the difference being that in the latter case, the pure state used in the construction is the maximally entangled Bell state. When $\epsilon=1$, the state $\hat{\cal{W}}_{X}(\epsilon)$ is completely mixed; when $\epsilon=0$, the state is pure. Since, in general, $\hat{\cal{W}}_{X}(\epsilon)$ is mixed, it does not have a ket representation, such as in (\ref{E2}). However, when $\epsilon=0$, the state is given by (\ref{E2}), with $b_{12}, b_{21} =0$, and the value of the angle $\theta$ determines how entangled the pure state is.

\begin{figure}
  \includegraphics[width=\linewidth]{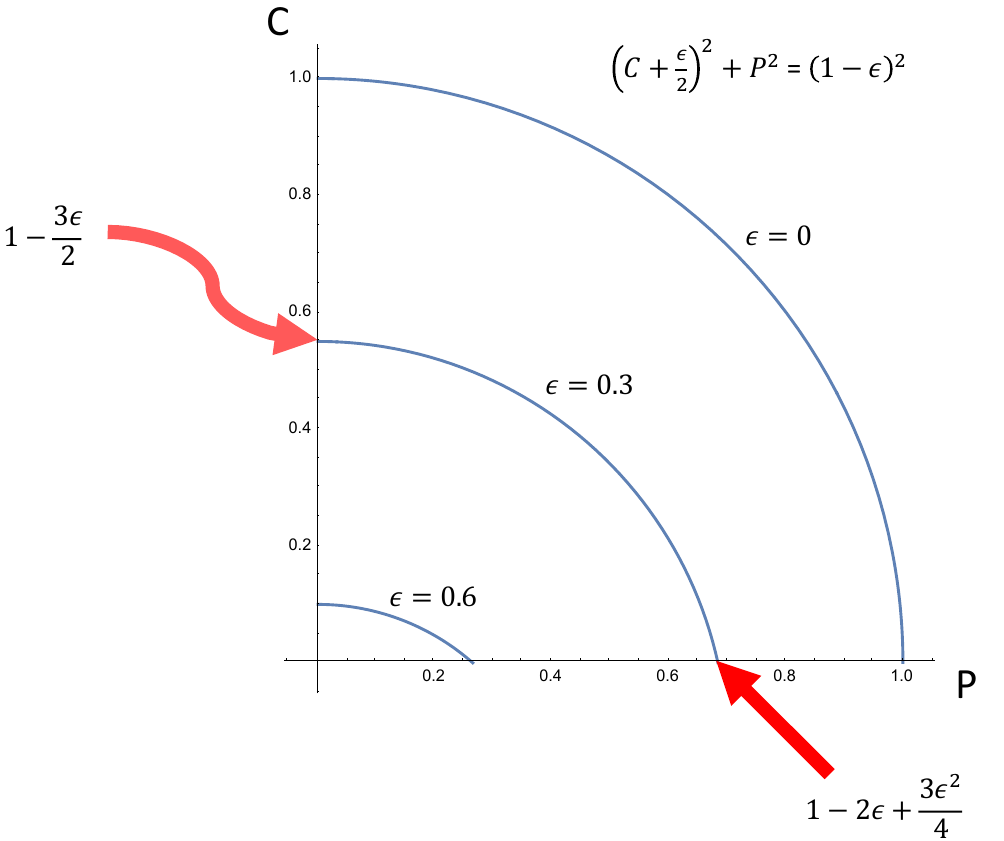}
  \caption{{\bf The Dependence of the Entanglement-Polarization relationship on mixedness.} The outermost curve, for $\epsilon=0$, corresponds to the pure state case, where the equality relation $C^2+P^2=1$ holds. Notice that as the level of mixedness increases, when $\epsilon$ increases in value, the circle that defines the $C$ and $P$ relationship shrinks. Finally, when $\epsilon$ decreases to $2/3$, the circle becomes a point in the origin. The red arrows give the general expressions for the value of $C$ and $P$ on their respective axes as functions of $\epsilon$.}
  \label{fig:figure2}
\end{figure}

The polarization and concurrence of $\hat{\cal{W}}_{X}(\epsilon)$ are given, respectively, by:
\begin{eqnarray}
P&=&(1-\epsilon) |\cos 2\theta\label{PMP}|\\
C &=& \max \left( 0, \mathscr{C} \right),\label{CMP}
\end{eqnarray}
where $\mathscr{C}=|(1-\epsilon)\sin 2 \theta|-\frac{\epsilon}{2}$.
Combining (\ref{PMP}) and  (\ref{CMP}) in the case when $C \ne 0$, we obtain:
\begin{equation}
\left(\mathscr{C}+\frac{\epsilon}{2}\right)^2+P^2=(1-\epsilon)^2.
\label{CPMP}
\end{equation}
The relationship is plotted in figure \ref{fig:figure2}. 

Figure \ref{fig:figure2} suggests an elegant geometrical interpretations of these results. We have seen in eq.(\ref{CPtheorem}) above that, for pure states, $C$ and $P$ can be thought of as the coordinates of a point on a unit circle centered at $(0,0)$. Our simple model of noise in such a system changes the radius of the circle to $1-\epsilon$ and breaks the symmetry between $C$ and $P$ by translating the center of the circle to $(0,-\epsilon/2)$. Let us think of the radius of the relationship between $C$ and $P$ as corresponding to an \emph{object}, such as an \emph{apple pie}. Although $C$ and $P$ remain a complementary pair in the mixed state case, the coherency \emph{pie} shrinks in size and shifts its center, so that they have less to share. In the extreme case, when the state is maximally mixed, the pie shrinks to a point, and there is nothing to share between the two quantities. The same effect is expected when a state is rendered mixed by ignoring other relevant DOFs describing the field (by tracing them out), as opposed to the addition of noise considered here.

%In conclusion, for classical light, the coherence in various DOFs depends on the entanglement between them. A coherence of particular interest, on the practical level, is polarization coherence. In previous work \cite{pureCP}, polarization (for a particular DOF) and entanglement (between all the DOFs) have been shown to be related by an equality when the state of the electric field is pure; i.e, when it can be described by a projector. Here, we looked at the more realistic case of adding phase noise to the system, making the electric field state lose its purity. 

\section{Conclusions}

In conclusion, in this work, we presented a model for a mixed state of classical light and studied how coherence relationships in the field description depends on the purity of the state. Specifically, we studied an experimental model in a Young interference set up that, with an optical manipulation, rendered the original pure state of the field into a mixed X-state. We demonstrated how the relationship between classical entanglement and polarization coherence for X-states is not, in general, given by the equality as in the pure-state case \cite{pureCP}. Moreover, we looked at a specific model of an X-state, that depends on one parameter (which determines its purity) and showed how the coherency pie shared between entanglement and polarization shrinks when the state is more mixed. One of the questions that arise is how the concept of duality, as discussed in \cite{Duality}, for example, is affected. Is there another property that is necessary to \emph{complete} the picture and get the equality relationship in the mixed-state case, or is it just a case where the total coherence in the system actually shrinks? In other words, is there another DOF that corresponds to a physical property in the system that will render it pure and satisfy the PCT equality or is there no such DOF, and, therefore, the mixed state cannot be purified physically (as opposed to formalistically), making it inevitable that the coherency pie is now smaller? 

This work was supported in part by the M. Hildred Blewett Fellowship of the American Physical Society, www.aps.org, by NSERC, and by NSF-INSPIRE GR 506175.
A. Al-Qasimi would like to thank J. H. Eberly, D. F. V. James, A. T. Friberg, and A. M. Steinberg for valuable discussions, and T. A. V. James and A. E. M. James for their cooperation.

\noindent \emph{Disclosures}: The author declares no conflict of interest.

\end{document}